\documentclass[conference]{IEEEtran}
\IEEEoverridecommandlockouts
\usepackage{cite}
\usepackage{amsmath,amssymb,amsfonts}
\usepackage[ruled,linesnumbered]{algorithm2e}
\usepackage{algpseudocode}
\usepackage{graphicx}
\usepackage{textcomp}
\usepackage{xcolor}
\usepackage{listings}
\usepackage{booktabs}
\usepackage{multirow}
\usepackage{subcaption}
\usepackage{lipsum}
\usepackage{calc}
\usepackage{soul}

\def\BibTeX{{\rm B\kern-.05em{\sc i\kern-.025em b}\kern-.08em
    T\kern-.1667em\lower.7ex\hbox{E}\kern-.125emX}}
\begin{document}

\RestyleAlgo{ruled}

\title{Adaptive Testing for LLM-Based Applications:\\ A Diversity-based Approach}

\author{
\IEEEauthorblockN{Juyeon Yoon}
\IEEEauthorblockA{\textit{School of Computing} \\
\textit{KAIST}\\
Daejeon, Republic of Korea \\
juyeon.yoon@kaist.ac.kr}
\and
\IEEEauthorblockN{Robert Feldt}
\IEEEauthorblockA{\textit{Dept. of Computer Science \& Engineering} \\
\textit{Chalmers University}\\
Gothenburg, Sweden \\
robert.feldt@chalmers.se}
\and
\IEEEauthorblockN{Shin Yoo}
\IEEEauthorblockA{\textit{School of Computing} \\
\textit{KAIST}\\
Daejeon, Republic of Korea \\
shin.yoo@kaist.ac.kr}
}

\newcommand{\fixme}[2][red]{\textcolor{#1}{FIXME: #2}}
\newcommand{\addcite}[2][orange]{\textcolor{#1}{ADDCITE: #2}}

\newcommand{\inlinecode}[1]{\lstinline[basicstyle={\ttfamily\small}]{#1}}
\newcommand{\smallinlinecode}[1]{\lstinline[basicstyle={\ttfamily\tiny}]{#1}}

\lstset{
  breaklines=true,
  xleftmargin=5pt,
  xrightmargin=5pt,
  aboveskip=10pt,
  belowskip=10pt,
  breakautoindent=false,
  breakindent=0ex,
  basicstyle=\scriptsize\ttfamily,
  backgroundcolor=\color{white},
  showstringspaces=false,
  frame=ltrb,
  language={},
  tabsize=2,
  numbers=none,
  keywordstyle=\color{shpurple}\textbf,
  commentstyle=\color{shgreen}\textit,
  stringstyle=\color{shred}
}

\newcommand\myhl[1]{{\sethlcolor{lightgray}\hl{#1}}}
\definecolor{lightergray}{gray}{0.9}
\newcommand\myhll[1]{{\sethlcolor{lightergray}\hl{#1}}}

\maketitle

\begin{abstract}
The recent surge of building software systems powered by Large Language Models (LLMs) has led to the development of various testing frameworks, primarily focused on treating \textit{prompt templates} as the unit of testing. Despite the significant costs associated with test input execution and output assessment, the curation of optimized test suites is yet overlooked in these tools, which calls for tailored test selection or prioritization strategies. In this paper, we show that diversity-based testing techniques, such as Adaptive Random Testing (ART) with appropriate string distance metrics, can be effectively applied to the testing of prompt templates. Our proposed adaptive testing approach adjusts the conventional ART process to this context by selecting new test inputs based on scores derived from existing test suite and their labelling results. Our results, obtained using various implementations that explore several string-based distances, confirm that our approach enables the discovery of failures with reduced testing budgets and promotes the generation of more varied outputs.
\end{abstract}


\section{Introduction}

The rapid advancements in Large Language Models (LLMs) have sparked widespread interest in integrating these models into software systems. These systems span a range of domains, including search engines~\cite{googleAIoverview, perplexityAI}, language education platforms~\cite{duolingoMax}, text-based games~\cite{yan2023larp, gptrpg, AIdungeon}, and coding assistants~\cite{githubCopilot}. Despite their potential and growing popularity, developing and testing LLM-based software presents significant challenges~\cite{chen2024empirical}. A critical difficulty lies in the inherent unpredictability of LLM-generated outputs, which are highly non-deterministic and thus challenging to control. A recent study of open-source LLM projects has highlighted poor quality and incorrect answers from LLMs as key concerns~\cite{cai_demystifying_2024}. Achieving the desired level of performance in these applications requires iterative and labor-intensive prompt engineering, as the developers need to continuously refine their prompts to guide the LLMs towards generating the desired outputs.

A typical prompt contains both a fixed part that describes the given task and a variable part that needs to be adapted to dynamic program contexts, such as user inputs or changing environments or changing state(s). To cope with this variability, a foundational component for handling queries to the LLM is constructed as ``prompt templates''~\cite{langchain_prompt_template}. Prompt templates combine natural language instructions to guide LLM generation with placeholders for context-specific inputs. However, optimising these templates becomes increasingly complex as developers need to ensure that they perform well across a wide variety of input contexts. Exhaustively testing all available inputs is infeasible due to the infinite variability, as well as high cost of executing LLM queries and manual effort required to analyse their outputs~\cite{shankar_spade_2024, shankar_who_2024}. Moreover, the iterative nature of prompt refinement necessitates efficient testing processes.

Given that most high-performance LLMs are closed-source and accessible only through remote APIs, a black-box testing strategy is a natural choice. In particular, diversity-based techniques such as Adaptive Random Testing (ART)~\cite{chen_adaptive_2005}, which strategically selects diverse inputs to ensure an even distribution across the input space, can be applied to any input data type including text. Besides the fact that it operates independently of internal program states, ART is well-suited for LLM applications as it has recently been shown to incur minimal overhead especially when the target program involves non-trivial execution times~\cite{biagiola_adaptive_2024}.

Building on these insights, we propose a test prioritization and selection method for prompt templates in LLM applications, inspired by ART. Similar to conventional ART, our approach iteratively selects the next test input that is farthest from the reference set of previously executed inputs. However, it extends this process by incorporating the outcomes of executed inputs to dynamically adjust the reference set used for distance calculation.

Our empirical evaluation on 46 prompt templates shows that diversity-based adaptive testing can efficiently select meaningful test inputs and uncover more failures within constrained testing budgets. Among the distance functions explored for diversity-based test prioritization, the Normalized Compression Distance (NCD)~\cite{cilibrasi2005clustering} shows the most promising results. It improves the average percentage of failure detection (APFD) by 7.24\% on average, with gains reaching up to 34.3\% compared to the random baseline. Furthermore, it produces outputs containing 9.5\% more unique words on average. Additionally, we observe that the effectiveness of distance metrics varies significantly across tasks and input distributions, underscoring the need for further investigation to optimize for specific testing scenarios.

Our approach complements existing early-stage testing frameworks for LLM applications by serving as a post-processing method to prioritize or select test inputs.
Existing techniques either focus solely on automating the benchmarking of prompt-templates~\cite{hudson_software_2024} or synthesize additional input data based on prompt contents and related information sources~\cite{deepeval_dataset_generation, promptfoo_dataset_generation}. An effective test selection and prioritization technique can reduce the cost testing, while improving the operational efficiency.

Our contributions are as follows:
\begin{itemize}
    \item We demonstrate the applicability of diversity-based black-box testing techniques, particularly ART, to LLM applications.
    \item We suggest an adaptive testing framework that can be extended to general scoring functions based on existing test suite and outcomes.
    \item We empirically evaluate various syntactic and semantic distance functions for input selection and prioritization, providing insights into their task-specific effectiveness.
\end{itemize}

\section{Background}

In this section, we provide an overview of the current state of LLM-based applications and testing practices, as well as black-box diversity-based testing techniques that can be applied to LLM applications.

\subsection{LLM-based Applications and Testing in Practice}

In LLM-based applications, developers typically implement prompt templates which contain both the ``template'' part that is invariant across different uses of the application, and the ``input placeholder'' parts that are filled with varying input values. Let us consider a simple RSVP email generator instantiated with the following prompt template, where the placeholder parts are enclosed in curly braces:

\begin{lstlisting}[escapechar=!]
Act as a skillful email generator. Write an RSVP email focusing in response to the following invitation email:
===
!\myhl{\{invitation\_email\}}!
===
The sender's intention about attendance is: !\myhl{\{intention\}}!. Include the following personal message from the sender in the email: !\myhl{\{personalization\}}!. Only generate the email text without any additional explanations.
\end{lstlisting}

Here, the prompt template can be treated as a standard function accepting a set of input variables: ``invitation\_email'', ``intention'', ``personalization'', and produces a string as output. Several specialized testing frameworks are available to evaluate these prompt templates~\cite{promptfoo, TruLens-Eval, langsmith, giskard, deepeval, arawjo_chainforge_2024}. They typically enable users to execute combinations of input variables and provide (visual) tools to compare outputs across various models and prompt configurations. An example test case for the above prompt template is as follows:

\begin{lstlisting}
- invitation_email: "Dear Sophia, we are thrilled to invite you to our Wedding Ceremony on June 1, 2025, at 3 PM at the Garden Pavilion. It would mean the world to us if you could celebrate our special day with us."
- intention: Accepting the invitation.
- personalization: "I am honored to join you on your special day and wish you both a lifetime of happiness."
  
\end{lstlisting}

Subsequently, the test suite would be collection of such test cases.

While the generation of such test suites is typically performed manually, some tools experimentally support automated dataset generation using generative models~\cite{deepeval_dataset_generation, promptfoo_dataset_generation}. These tools synthesize input sets for testing by leveraging prompt content or related information as context. However, they do not guarantee the ``validity'' of the generated inputs (i.e., their appropriateness for the target prompt) or their diversity (i.e., their coverage of the input space). This limitation can result in irrelevant or redundant tests, increasing both cost and complexity. We advocate for a systematic method to construct optimized test suites that increase both efficiency and effectiveness.

\subsection{Diversity-based Testing}

Diversity among test inputs is essential in software testing to ensure the overall quality of the target system. Adaptive Random Testing (ART) is a representative diversity-based approach, proposed as an improvement over basic random testing~\cite{Chen:2005qy}.
ART selects the input farthest from previously executed inputs among a set of randomly sampled candidates, enabling broader exploration of the input space. This process typically involves calculating pairwise distances between executed inputs and candidate samples at each step.

Although ART is conceptually simple and has demonstrated effectiveness over random testing, it is computationally expensive due to the quadratic complexity of distance calculations relative to the number of inputs~\cite{arcuri2011adaptive}. However, studies suggest that when test execution times are sufficiently high, the cost of these calculations becomes negligible, as they can be performed in parallel with test execution. This is particularly relevant for testing LLM-based applications, where query execution times are often significant, and the stochastic behavior of LLMs necessitates multiple executions of the same input to evaluate output consistency.
Recent studies have also proposed other ways to speed up distance calculations in ART~\cite{biagiola_adaptive_2024}.

Another diversity-based test selection approach leverages the concept of test set diameter (TSDm)~\cite{feldt_test_2016}. By extending pairwise normalized compression distance (NCD) to a multiset setting, TSDm both quantifies the diversity of test sets and enables the selection of diverse subsets from an initial pool. Empirical studies on diversity-based test prioritization~\cite{henard2016comparing} have shown that TSDm achieves superior fault detection rates compared to other black-box techniques. However, TSDm's practical adoption is often limited by its high computational cost, which exceeds even that of ART, as it scales quadratically with the size of the initial test input pool.

\section{Approach}

Our study focuses on a test selection or prioritization scenario, where a large set of initial test inputs, consisting of collected user data or synthesized inputs for a prompt template, is available for refinement. In this section, we present our approach, which adapts the original ART procedure to this context.

\subsection{Adaptive test selection method for prompt templates}
We propose a black-box test selection and prioritization method, detailed in Algorithm~\ref{alg:divtesting_framework}, which adaptively selects new test inputs based on previously selected ones. While inspired by Adaptive Random Testing (ART), our approach is specifically adapted to the test selection and prioritization of prompt templates. The original ART algorithm~\cite{chen_adaptive_2005} was designed for use with a random input generator that uniformly samples the input space, making it not directly applicable to our context. However, the core idea of ART, selecting the next test to be farthest from the already selected ones, remains relevant. Prior study~\cite{Leon:2003vn} suggest that reducing redundancy among tests, thus diversifying the test suite subset, can be efficient in revealing defects in test prioritization scenario.

Our algorithm selects candidates from the existing test pool (line 4) instead of generating new inputs. For each candidate, it computes a score based on its distance from the already selected tests and selects the candidate with the highest score (line 5). The chosen candidate is then removed from the pool, added to the test suite ($selectedTests$) (line 6-7), and executed on the target prompt template using the base model under test (PUT) (line 8). This process continues until the desired number of tests, $N$, is reached.

The algorithm supports both selection and prioritization scenarios. By setting 
$N$ to the size of the initial pool, it determines and prioritizes the execution order for all inputs.

\begin{algorithm}
\caption{Adaptive test selection and prioritization.}
\label{alg:divtesting_framework}
\SetAlgoLined
\KwIn{An initial pool of candidate tests $initialTests$, a target suite size $N$, a specified prompt template and a base model under test $PUT$.}
\KwOut{A resulting test suite $selectedTests$ of size $N$, adaptively selected from the initial pool.}

$pool \gets$ $initialTests$.\text{copy}()\;
$selectedTests \gets \emptyset$\;
\While{$|selectedTests| < N$}{
    $Cands \gets pool.\text{sample}()$\;
    $bestCand \gets \arg\max_{c \in Cands} \text{calculate\_score}(c, selectedTests)$\;
    $pool.\text{remove}(bestCand)$\;
    $selectedTests.\text{add}(bestCand)$\;
    $PUT.\text{execute}(bestCand)$\;
}
\Return $selectedTests$\;
\end{algorithm}

\subsection{Diversity-based score calculation w/ selective reference set}
\label{sec:approach_score_calculation}

The scoring function, \inlinecode{calculate_score}, determines the best candidate to be selected at each iteration of the Algorithm~\ref{alg:divtesting_framework}. We consider diversity as the primary criterion for scoring; Algorithm~\ref{alg:divtesting_score} defines a diversity-based scoring function used in our study. Similar to the standard ART~\cite{chen2010adaptive}, this function calculates the score of the candidate input by using the minimum distance between the candidate and each individual test in the existing test suite. 
We explore various implementations using different string-based distances, given that prompt template inputs are typically textual. We also note that the scoring function is pivotal to the flexibility of our approach; our suggested method can be generalized to various scoring functions as far as they maintain the intuitive property of giving higher scores to the candidate inputs that add more ``meaningful'' information to the existing test suite.

Additionally, our method enables selectively filtering the previously executed tests based on specific criteria (line 3). The \inlinecode{select\_references} function constructs a set of tests used to calculate the distance-based scores, which we refer to as the \textit{reference set}, by subsetting the entire set of executed tests ($selectedTests$).

\begin{algorithm}
    \caption{Diversity-based score calculation}
    \label{alg:divtesting_score}
    \SetKwFunction{FcalculateScore}{calculate\_score}
    \SetKwProg{Fn}{Function}{:}{}
    \Fn{\FcalculateScore{$c$, $selectedTests$}}{
        $score \gets \infty$\;
        $referenceSet \gets \text{select\_references}(selectedTests)$\;
        \ForEach{$t \in referenceSet$}{
            $d \gets distance(c, t)$\;
            \If{$d < score$}{
                $score \gets d$
            }
        }
        \Return $score$
    }
    \end{algorithm}

\begin{figure}[t]
    \centering
    \includegraphics[width=0.9\linewidth]{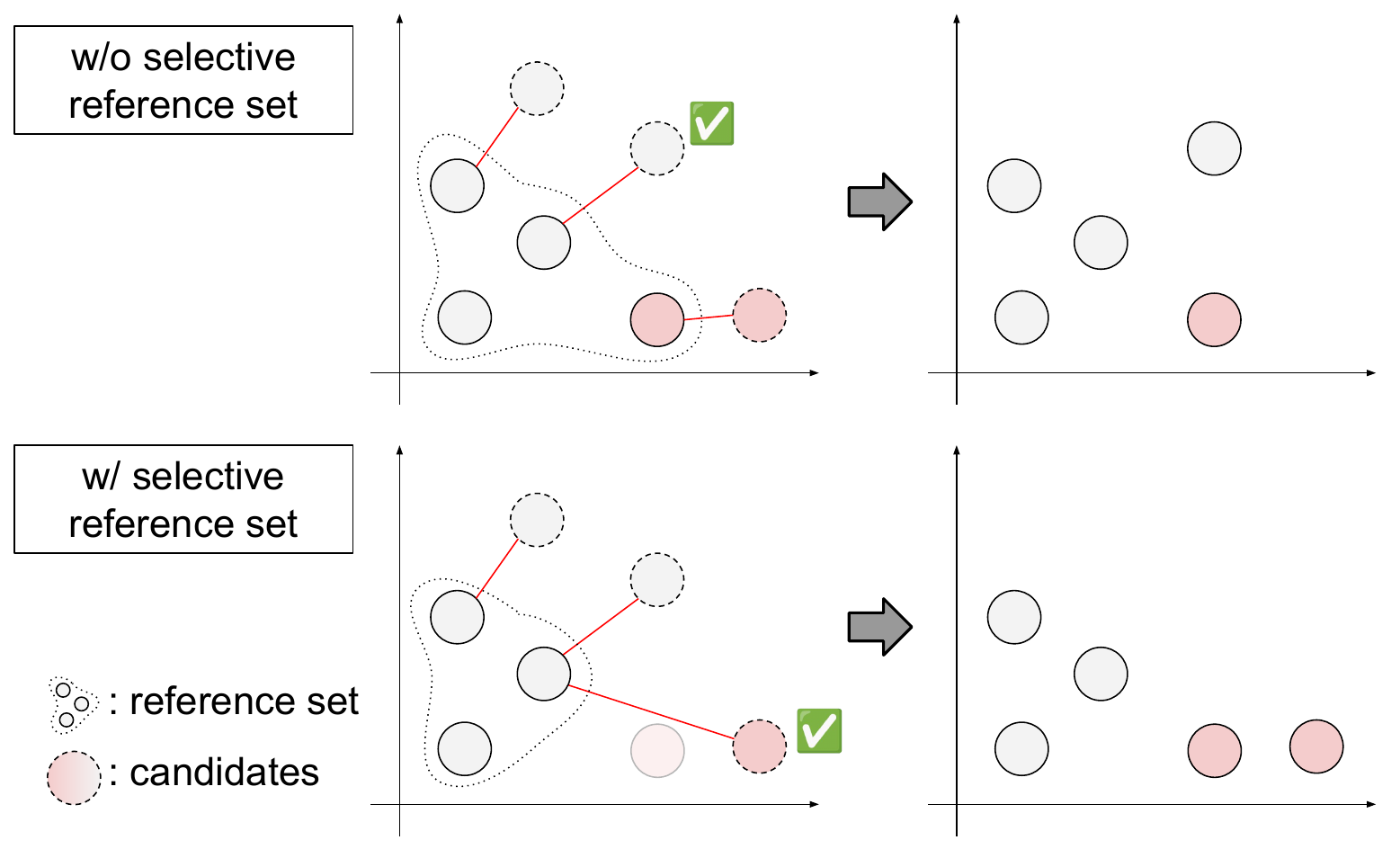}
    \caption{Adaptive testing framework for prompt templates.}
    \label{fig:selective_reference_set}
\end{figure}

ART assumes the existence of contiguous failure and non-failure regions in the input space~\cite{chen2010adaptive}, implying that selecting inputs far from existing ones increases the likelyhood of discovering new faults. Standard ART uses all executed inputs, both failing and passing, as the reference set. By design, similar failures are treated as an indication of the same fault, thus identifying multiple relevant failures is not desired. However, in LLM applications, where ``faults'' within natural language prompts lack precise definitions, we posit that identifying multiple similar failures can be beneficial to developers. For example, developers can find a recurring pattern of incorrect outputs, and adjust the prompt template accordingly. We suggest a selective reference set strategy upon this intuition, incorporating a subset of ``passing'' inputs from the executed tests. With this modification, we expect to increase the likelihood of failure detection while acknowledging that parts of the detected failures may be similar. Figure~\ref{fig:selective_reference_set} illustrates a possible case that the selective reference set strategy can help incorporate more failing inputs.

To filter out the failing inputs, we require a specific criteria for passing and failing inputs; however, the stochastic nature of LLMs can produce different outputs for the same input across multiple executions, making the determination of passing/failing inputs non-trivial. We address this ambiguity by deciding the correctness of an output based on the ground-truth expected output, and assign the pass/fail status of an input based on the correctness ratio over multiple executions. Formally, let \( T \) be the set of all executed tests, and \( \text{exec}(t) \) denote the set of outputs produced by \( t \) after \( n \) executions. Let \( O_{\text{expected}}(t) \) be the ground-truth output for test \( t \). We use the the correctness function for an individual execution result defined as:
\[
\text{isCorrect}(o, O_{\text{expected}}(t)) =
\begin{cases} 
1 & \text{if } o \text{ matches } O_{\text{expected}}(t), \\
0 & \text{otherwise}.
\end{cases}
\]

For a test \( t \), we compute the correctness ratio over \( n \) executions:
\[
\text{correctnessRatio}(t) = \frac{\sum_{o \in \text{exec}(t)} \text{isCorrect}(o, O_{\text{expected}}(t))}{n}.
\]

Defining the threshold for majority correctness as \( \tau \) (e.g., \( \tau = 0.5 \) in our experiments), the selective reference set \( \text{select\_references}(T) \) can now be expressed as:
\[
\text{select\_references}(T) = \{ t \in T \mid \text{correctnessRatio}(t) \geq \tau \}.
\]

\section{Experimental Setup}

This section provides details about experimental setup.

\subsection{Research Questions}
Our evaluation aims to answer the following questions.
\subsubsection{RQ1. Failure Discovery} How does the diversity-based adaptive testing improve failure discovery for LLM applications? Specifically, we explore the two sub-questions:
\begin{itemize}
    \item RQ1-1. How do different distance metrics for implementing the proposed test selection/prioritization method perform in increasing the rate of failure detection with fewer test inputs?
    \item RQ1-2. How does the selective reference set strategy affect the failure discovery rate?
\end{itemize}

\subsubsection{RQ2. Output Diversity} To what extent does the diversity-based adaptive testing approach promote the generation of more varied outputs?

\subsubsection{RQ3. Cost Analysis} What is the computational overhead of selecting new test inputs using various distance metrics, and how does employing a selective reference set strategy affect this cost?

\subsection{Dataset}
To evaluate our diversity-based test selection and prioritization method, we constructed a dataset of 46 prompt templates sourced from two LLM evaluation datasets: BIG-Bench Hard (BBH)~\cite{bbh-dataset} and Public Pool of Prompts (P3)~\cite{p3-dataset}. These prompts cover a diverse range of tasks, including arithmetic, logical reasoning, and language understanding. Crucially, these datasets provide fixed templates for prompts along with input/output examples for each task, enabling the construction of an initial test suite and automating output correctness assessment. 

The prompt templates include a standardized instruction that constrains outputs to a specific format, such as \inlinecode{Provide the final answer in the format of "The answer is [answer]."}. By checking whether the generated output contains the expected answer, the output evaluation can be automated.

\subsection{Models and Metrics}

The effectiveness of a test suite is typically measured by structural coverage metrics or fault detection capability. However, since our evaluation focuses on prompt templates tested against closed-source LLMs (GPT-4o for P3 dataset and GPT-4o-mini for BBH dataset in our experiments), traditional coverage metrics are not directly applicable. Instead, we focus on failure detection capability, emphasizing the importance of identifying incorrect outputs in LLM-based applications (see Section~\ref{sec:approach_score_calculation} for detailed rationale). 
To this end, we adopt the average percentage of failure detection (APFD), a straightforward adaptation of the Average Percentage of Faults Detected metric~\cite{do2006use}, which has been widely used for evaluating test prioritization techniques. APFD, which ranges from 0 to 100, indicates the rate of failure detection, with higher values reflecting faster identification of failures. 
Additionally, we assess output diversity by calculating the average number of unique words in the outputs generated from each input subset. Prior studies~\cite{alshahwan2012augmenting} suggest that output diversity correlates with fault-finding capability, making it a valuable proxy for evaluating the quality of a test suite.

\subsection{Baseline and Comparison Targets}
We use random selection as a baseline for comparison, and additionally incorporate TSDm with NCD multiset extension~\cite{feldt_test_2016} as an additional diversity-based selection strategy.

The distance metrics used for implementing various instances of our test selection method are as follows:
\begin{itemize}
    \item Normalized Compression Distance (NCD)~\cite{cilibrasi2005clustering}
    \item Cosine distance w/ 2-gram embeddings
    \item Cosine distance w/ Sentence-BERT~\cite{reimers2019sentence} embeddings
\end{itemize}

As our approach, with the diversity-based scoring function, is strongly inspired by ART, we refer to our prioritization and selection method described in Section~\ref{sec:approach_score_calculation} as ART to present our results for simplicity; we refer to the different implementations of the proposed diversity-based adaptive testing method as ART\_NCD, ART\_2gram, and ART\_sBERT, respectively.

For random and the ART variants, we repeat the experiments 100 times for BBH and 10 times for P3, adjusting for the larger test input pool in P3 (up to 1,000 vs. 100 for BBH).

\section{Experimental Results}

This section provides our experimental results.

\subsection{Failure Discovery (RQ1)}
To answer RQ1, we evaluate the percentage of failures revealed by selected test inputs with various selection strategies.

\subsubsection{RQ1-1. Selection methods and failure discovery}

\begin{figure}[t]
    \centering
    \begin{subfigure}[b]{\columnwidth}
        \centering
        \includegraphics[width=\textwidth]{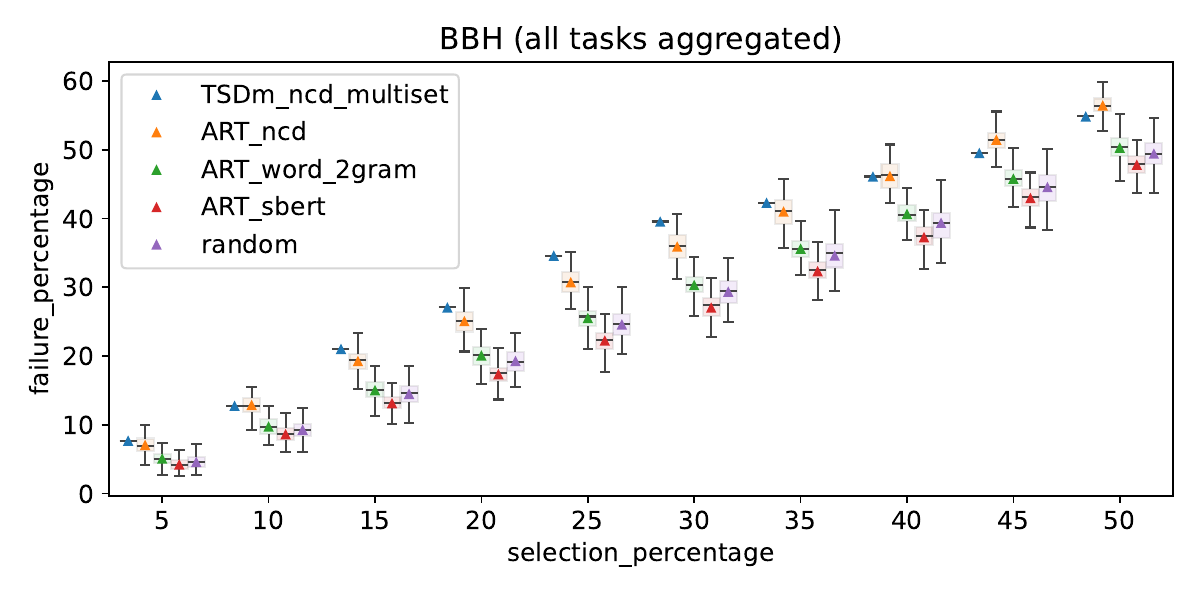} 
        \caption{BBH Dataset}
        \label{fig:RQ1_failure_percentage_bbh}
    \end{subfigure}
    \hfill
    \begin{subfigure}[b]{\columnwidth}
        \centering
        \includegraphics[width=\textwidth]{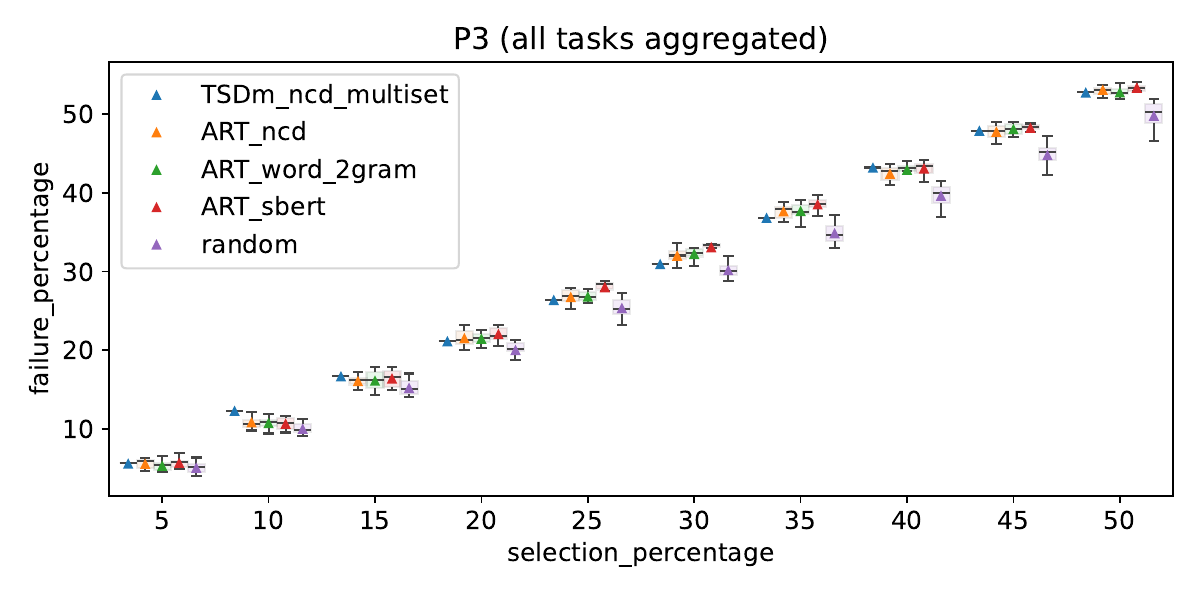} 
        \caption{P3 Dataset}
        \label{fig:RQ1_failure_percentage_p3}
    \end{subfigure}

    \caption{Percentage of found failures by varying selection percentages with different selection methods.}
    \label{fig:RQ1_failure_percentage}
\end{figure}

\begin{table}[ht]
    \centering
    \caption{Comparison of APFD for all tasks.}
    \label{tab:taskwise_APFD}
    \scalebox{0.9}{
    \begin{tabular}{@{}c|l|lllll@{}}
    \toprule
    & & \multicolumn{5}{c}{APFD} \\
    \midrule
    SRC & Task & TSDm & NCD & 2gram & sBERT & random \\
    \midrule
    \multirow{26}{*}{\textbf{BBH}}
    & bool\_exp & \textbf{53.7} & 50.0 & 48.4 & 45.0 & 50.3 \\
    & causal & 51.6 & 51.3 & \textbf{52.1} & 52.1 & 49.9 \\
    & date & \textbf{59.9} & 50.2 & 51.1 & 52.5 & 50.4 \\
    & disambig\_qa & \textbf{55.1} & 50.8 & 51.4 & 53.3 & 50.3 \\
    & dyck\_lang & \textbf{71.6} & 71.3 & 62.7 & 51.6 & 49.8 \\
    & fallacies & 56.9 & \textbf{60.4} & 49.4 & 49.7 & 49.4 \\
    & geometric & 63.4 & \textbf{64.9} & 62.8 & 39.1 & 49.2 \\
    & hyperbaton & 49.9 & \textbf{64.6} & 56.3 & 36.5 & 48.8 \\
    & deduction\_3 & 37.2 & 35.6 & 45.7 & 45.5 & \textbf{49.9} \\
    & deduction\_5 & \textbf{49.8} & 49.4 & 45.7 & 47.5 & 49.3 \\
    & deduction\_7 & \textbf{61.4} & 55.4 & 46.5 & 49.7 & 49.4 \\
    & movie & 46.6 & 51.1 & 51.5 & \textbf{51.7} & 49.8 \\
    & arithmetic & 60.7 & 53.2 & 46.8 & \textbf{61.6} & 51.0 \\
    & navigate & 75.8 & \textbf{82.6} & 67.3 & 48.7 & 48.3 \\
    & object\_counting & \textbf{80.8} & 60.3 & 51.7 & 26.1 & 50.6 \\
    & penguins & 36.5 & 46.3 & 48.8 & \textbf{54.9} & 50.5 \\
    & colored & \textbf{50.7} & 40.1 & 33.6 & 34.5 & 48.6 \\
    & ruin\_names & 43.0 & 45.9 & 44.9 & \textbf{51.6} & 50.5 \\
    & trans\_error & 49.1 & 48.9 & 49.6 & 47.6 & \textbf{50.1} \\
    & snarks & 57.3 & \textbf{58.3} & 55.6 & 52.6 & 50.3 \\
    & sports & \textbf{56.0} & 51.0 & 46.3 & 49.3 & 50.2 \\
    & temporal & 58.4 & \textbf{60.3} & 50.8 & 55.3 & 51.4 \\
    & tracking\_3 & 49.2 & 53.0 & 56.9 & \textbf{58.5} & 48.8 \\
    & tracking\_5 & 44.2 & 49.0 & 48.2 & 49.6 & \textbf{50.5} \\
    & tracking\_7 & 47.3 & \textbf{55.9} & 44.9 & 50.8 & 50.4 \\
    & word\_sorting & \textbf{61.4} & 61.2 & 61.4 & 46.6 & 50.0 \\
    \midrule
    \multirow{20}{*}{\textbf{P3}} 
    & news & 50.2 & 53.7 & \textbf{56.6} & 54.1 & 50.6 \\
    & arc\_challenge & 48.4 & 49.9 & 48.2 & \textbf{55.2} & 49.0 \\
    & amazon\_review & 44.7 & 45.9 & 56.1 & \textbf{59.7} & 52.2 \\
    & anli & 47.5 & 48.1 & 51.1 & \textbf{51.7} & 50.0 \\
    & dbpedia & 50.8 & 51.9 & 54.9 & \textbf{66.6} & 50.1 \\
    & dream & \textbf{56.6} & 50.4 & 49.6 & 50.7 & 48.7 \\
    & glue\_mrpc & 49.8 & 49.9 & \textbf{51.3} & 49.8 & 50.2 \\
    & hellaswag & 47.3 & 49.1 & 49.3 & \textbf{49.8} & 49.7 \\
    & paws & \textbf{52.7} & 50.6 & 48.0 & 49.7 & 48.7 \\
    & qasc\_qa & 69.6 & \textbf{70.6} & 48.0 & 53.9 & 47.0 \\
    & ropes & 45.2 & \textbf{59.5} & 52.6 & 57.5 & 48.9 \\
    & rotten\_tomatoes & \textbf{55.4} & 53.9 & 47.2 & 48.1 & 49.9 \\
    & social\_i\_qa & 46.7 & 47.1 & 49.5 & 48.0 & \textbf{50.5} \\
    & glue\_record & 50.2 & 48.9 & 48.5 & \textbf{50.3} & 49.6 \\
    & glue\_rte & 42.6 & 43.6 & 45.5 & 48.6 & \textbf{53.1} \\
    & glue\_wic & 49.7 & 47.7 & \textbf{51.8} & 48.4 & 49.9 \\
    & trec & \textbf{61.1} & 59.5 & 57.2 & 51.2 & 48.0 \\
    & wiki\_qa & 60.1 & \textbf{61.8} & 61.0 & 52.8 & 51.7 \\
    & wiqa\_effect & 50.2 & 49.6 & 49.7 & 49.7 & \textbf{50.5} \\
    & yelp\_review & 50.2 & 49.9 & \textbf{50.7} & 50.0 & 49.8 \\
    \bottomrule
    \end{tabular}}
    \end{table}

\begin{figure}[t]
    \centering
    \begin{subfigure}[b]{\columnwidth}
        \centering
        \includegraphics[width=\textwidth]{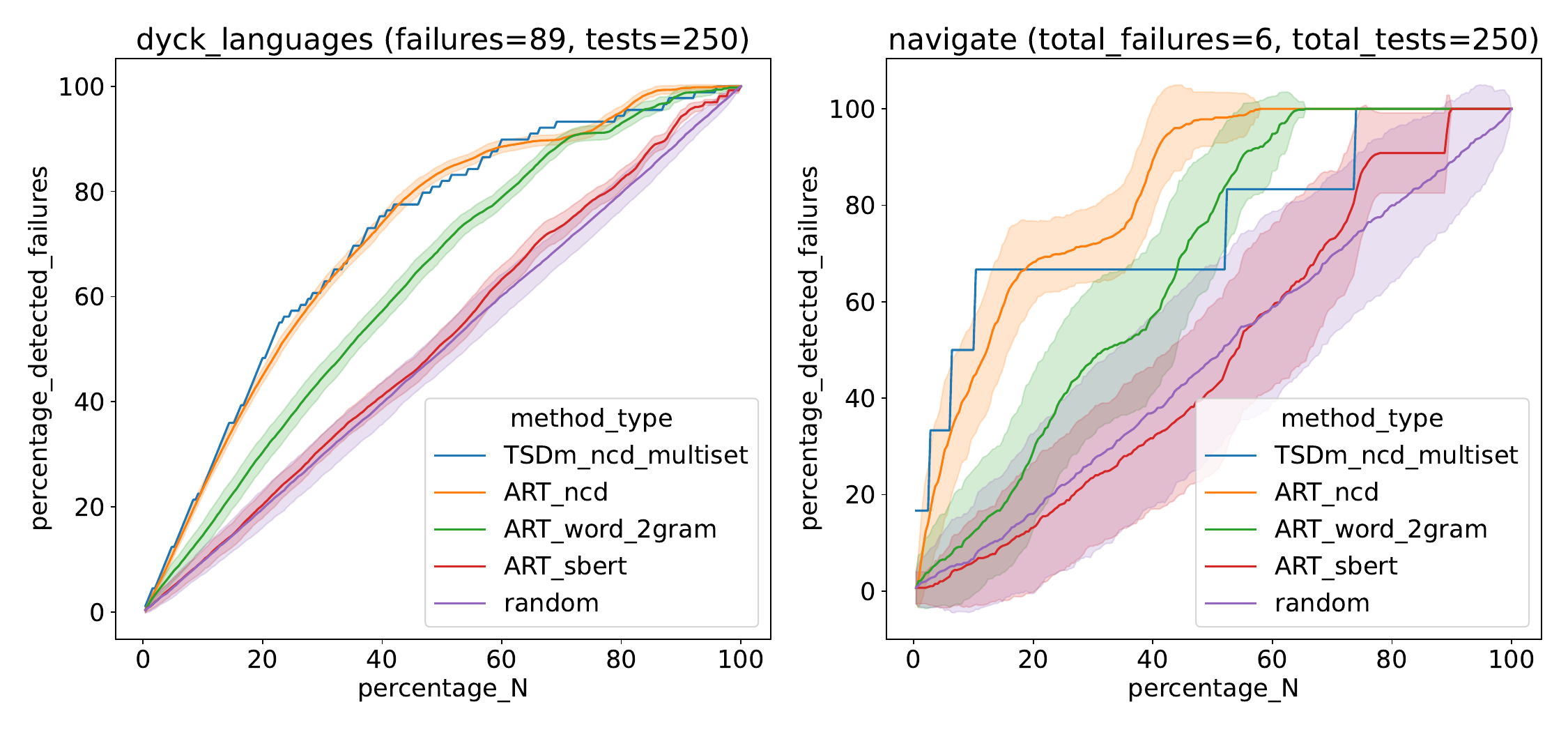} 
        \caption{Example tasks that NCD performs relatively well.}
        \label{fig:RQ1_NCD_tasks_APFD}
    \end{subfigure}
    \hfill
    \begin{subfigure}[b]{\columnwidth}
        \centering
        \includegraphics[width=\textwidth]{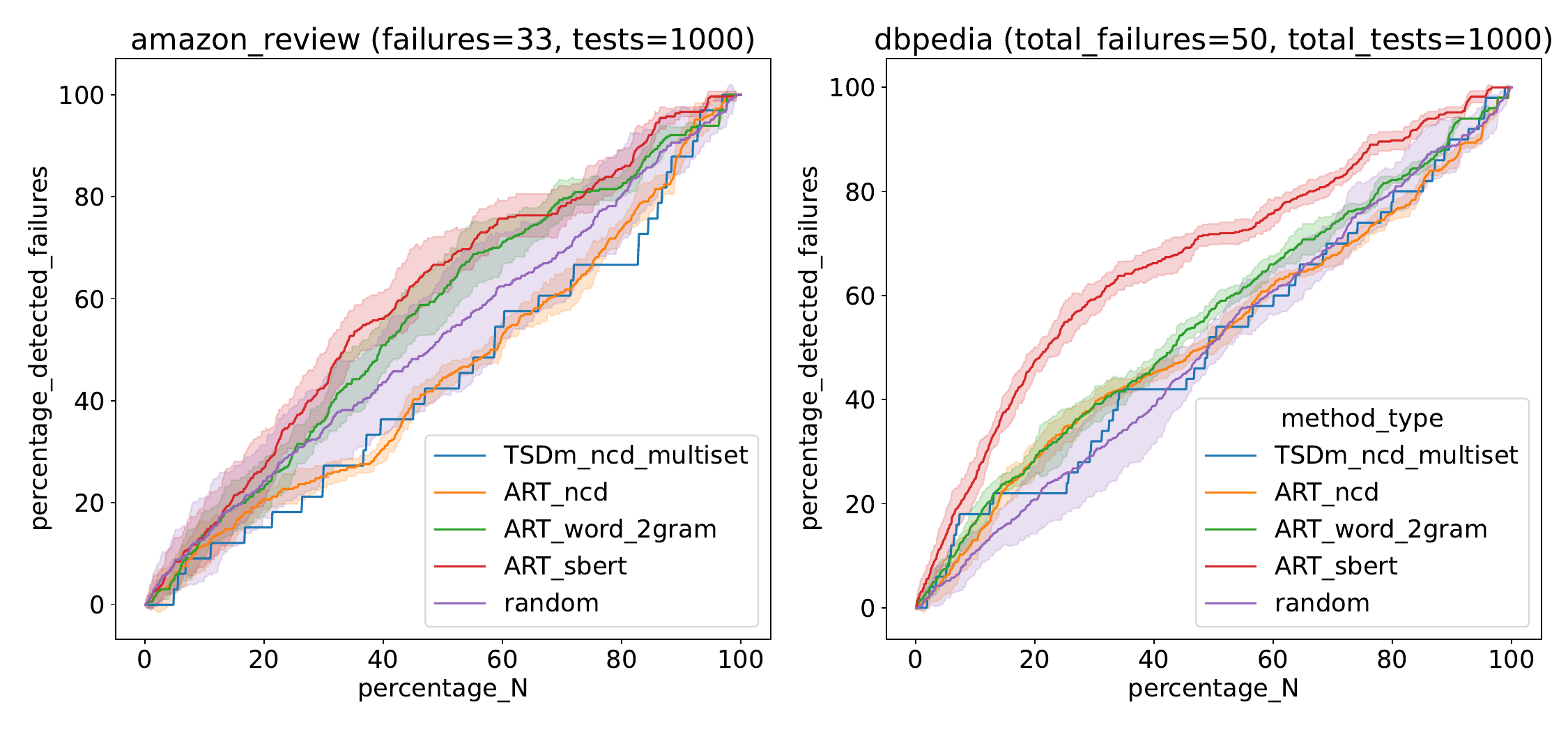} 
        \caption{Example tasks that sBERT-based distance performs relatively well.}
        \label{fig:RQ1_sBERT_tasks_APFD}
    \end{subfigure}

    \caption{Failure detection by the number of executed test inputs in test prioritization scenario.}
    \label{fig:RQ1_APFD_task_examples}
\end{figure}

\begin{figure}[ht]
    \centering
    \includegraphics[width=\linewidth]{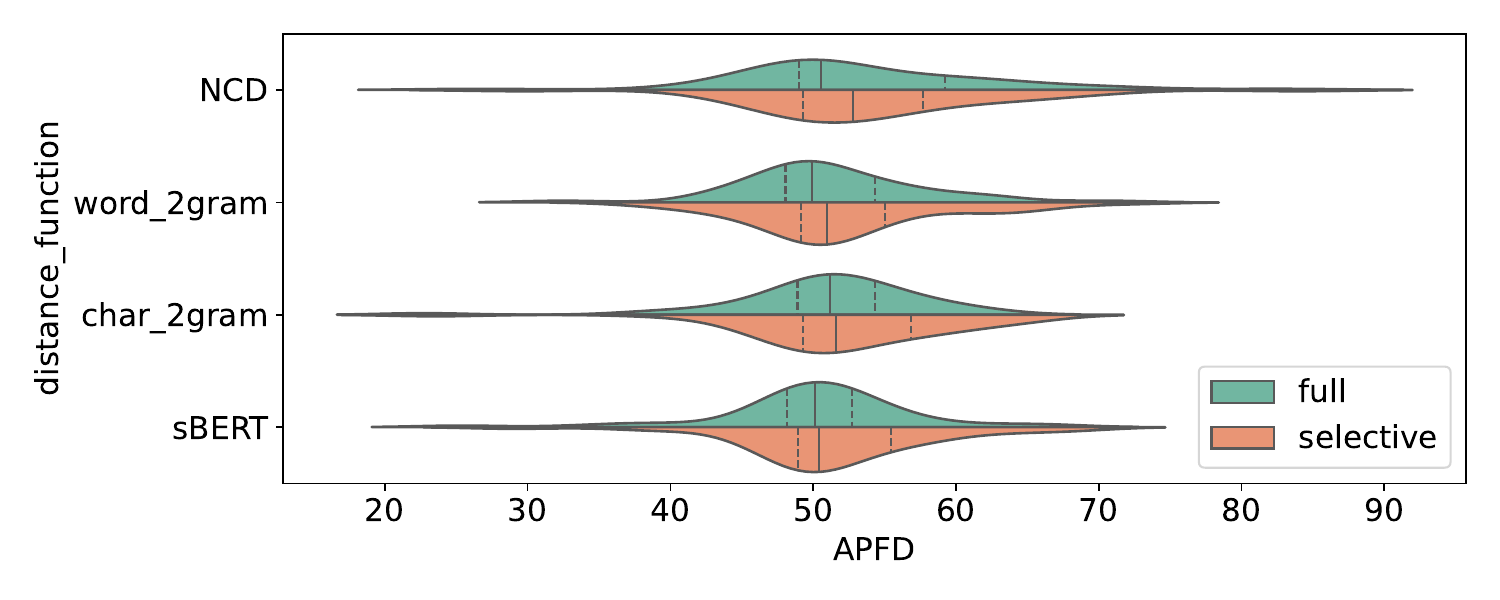}
    \caption{Distribution of APFD values with and without selective reference set selection.}
    \label{fig:RQ1_refset}
\end{figure}

First, we report the revealed failure percentages by different selection methods and selection percentages. Figure~\ref{fig:RQ1_failure_percentage} presents the average percentage of total failures across all tasks in both datasets. Overall, diversity-based selection methods, particularly NCD-based ones (TSDm and ART\_NCD) for BBH dataset and ART\_sBERT for P3 dataset, exhibit relatively higher failure percentages. 
Table~\ref{tab:taskwise_APFD} presents task-wise results in terms of APFD values. TSDm achieves the highest APFD values for most tasks (14 out of 46), improving the APFD values by 7.0\% on average (up to a maximum of 30.3\%); ART\_NCD (denoted simply as NCD in the table) also shows competitive performance, improving the APFD values by 7.24\% (34.3\%) compared to random selection. Wilcoxon Signed-Rank tests confirms that TSDm and ART\_NCD show statistically significant improvements in APFD values across the 46 tasks ($p = 0.014$ for TSDm, $p = 0.007$ for ART\_NCD). ART\_2gram and ART\_sBERT do not exhibit statistically significant improvements over random selection.

Although the overall results suggest the potential of diversity-based methods to improve failure detection, the performance of each method varies significantly across tasks and datasets. Figure~\ref{fig:RQ1_APFD_task_examples} illustrates the failure detection rate over the number of executed test inputs for several representative tasks. ART\_sBERT shows improvements in specific tasks, such as `penguins' task in BBH dataset, `amazon\_review' and `dbpedia' task in P3 dataset. On the other hand, NCD based methods dominate in certain tasks, such as `dyck\_lang', `navigate' and `object\_counting' tasks in BBH dataset. These tasks involve distinct challenges: `dyck\_lang' is about generating balanced paranthesis strings, `navigate' requires inferring destination positions after a series of instructions, and `object\_counting' requires counting the objects from given descriptions. The inputs for these tasks are often hard to represent as semantic embeddings (e.g., `dyck\_lang' inputs consist of parenthesis symbols), which may explain the better performance of NCD-based methods.
From this observation, one promising direction for future exploration would be predicting the most effective distance metric for a given task, based on task characteristics or the distribution of test inputs. We refer readers to Section~\ref{sec:discussion_distance_prediction} for further discussion.

\subsubsection{RQ1-2. Selective reference set strategy}

\begin{figure}[t]
    \centering
    \includegraphics[width=\columnwidth]{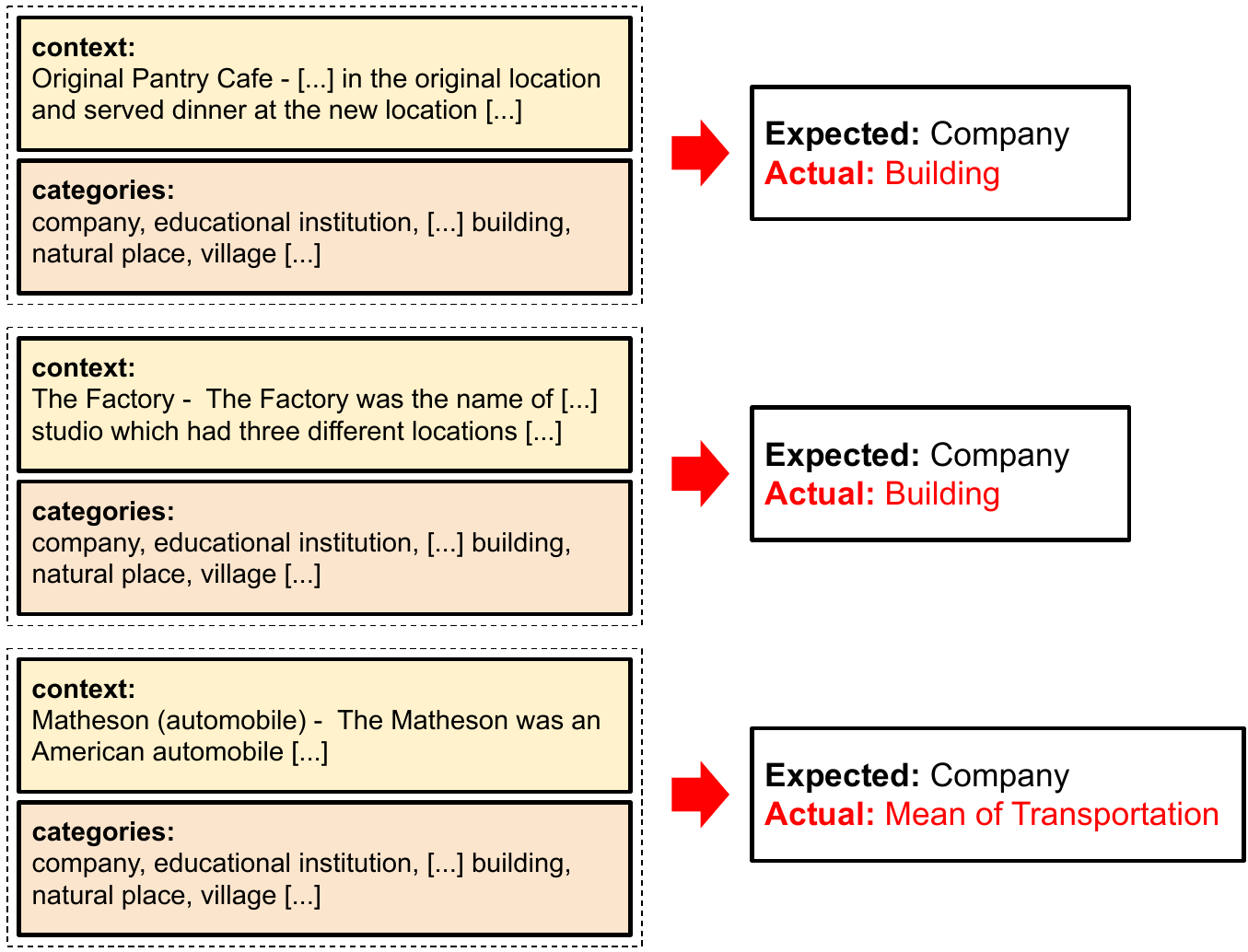}
    \caption{Example failing inputs contained from the `dbpedia' task in P3 dataset.}
    \label{fig:divtesting_failing_input}
\end{figure}

We also report the failure discovery results when the selective reference set strategy is applied. Figure~\ref{fig:RQ1_refset} illustrates the distribution of APFD values across all explored tasks, comparing ART with and without the selective reference set strategy. Across all distance metrics, the selective strategy consistently enhances APFD values. Specifically, ART\_NCD with selective reference set improves the APFDs in 32 out of 46 tasks, with an average increase of 0.96\% (up to a maximum of 9.32\%). Similarly, ART\_char\_2gram, ART\_word\_2gram and ART\_sBERT with selective strategy improves APFDs in 30, 33, and 28 tasks over the full reference set setting, respectively. The Wilcoxon Signed-Rank tests confirm these improvements for all distance metrics studied, with statistical sigificance ($p = 0.007$ for ART\_NCD, $p = 0.003$ for ART\_char\_2gram, $p = 0.003$ for ART\_word\_2gram, and $p = 0.005$ for ART\_sBERT).

Our underlying assumption of selective set strategy is that revealing multiple similar failures early on is advantageous, as they can guide developers toward prompt improvements. To hint at the validity of this assumption, we present example failing inputs from the `dbpedia' task illustrated in Figure~\ref{fig:divtesting_failing_input}. The task aims to classify a given context into a specific category, and requires two input variables: a target context to classify and a list of categories. The first and second ones refer to a similar issue; the LLM confuses the contexts, which describes companies located in various places, with a description of a specific building due to the location details contained in the context inputs. Although these inputs possibly require the same modification in the prompt, both failing inputs would be valuable in identifying a \textit{common} pattern of mistakes that LLM makes. Developers can use this insight to refine their prompt templates to avoid such mistakes, for example, by including few-shot examples that address the pattern or adding clarifications about category criteria (e.g., ``If the context mentions multiple locations but emphasizes operations, employees, or services, classify it as [Company]'').

\subsection{Output Diversity (RQ2)}

\begin{figure}[t]
    \centering
    \includegraphics[width=\columnwidth]{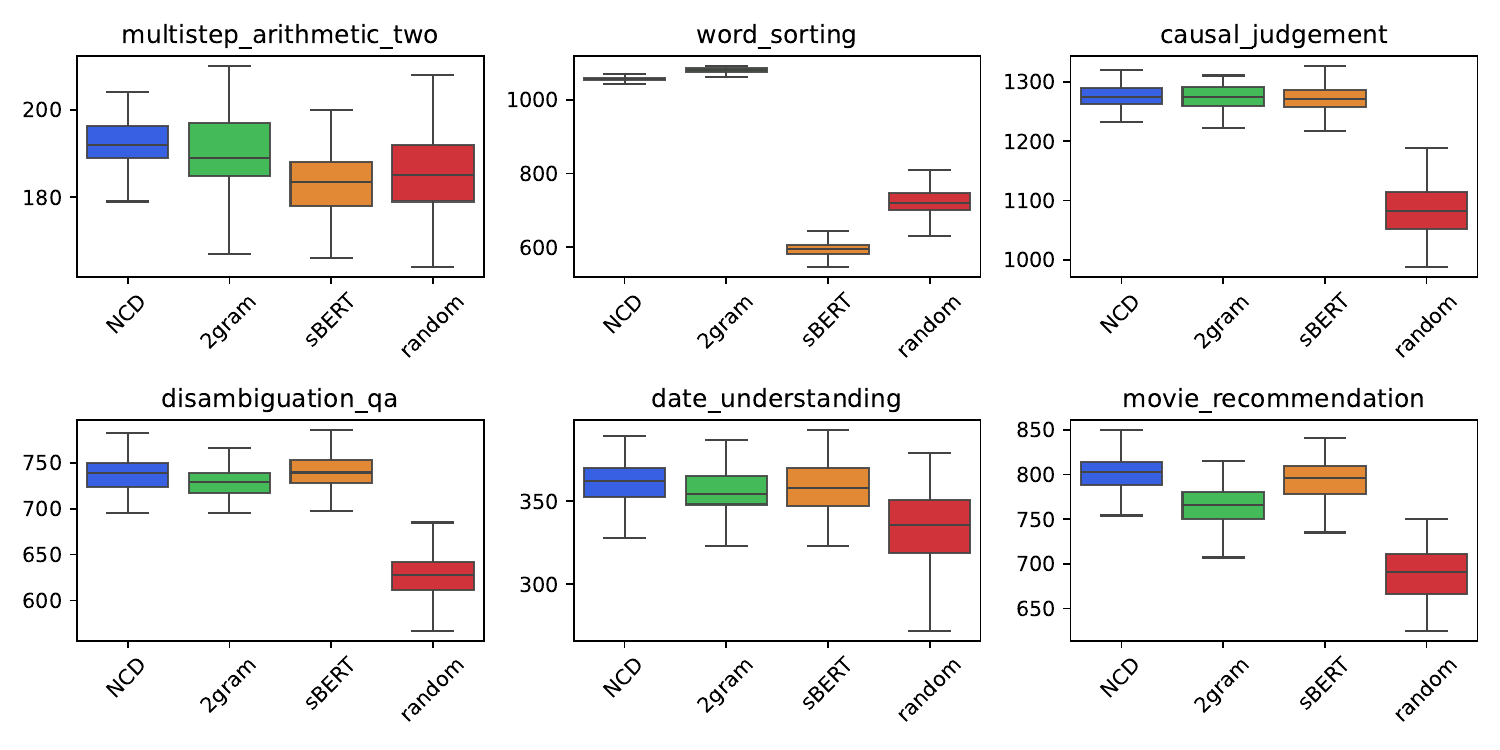} 

    \caption{The number of unique words contained in the outputs from $N$ selected test inputs. ($N$ = 50)}
    \label{fig:RQ2_output_diversity}
\end{figure}

To answer RQ2, we report the number of unique words contained in the generated outputs from input sets selected using studied techniques. Figure~\ref{fig:RQ2_output_diversity} presents the distribution of unique output word counts for randomly sampled six tasks. 

Across all tasks, outputs from ART\_NCD show a 9.5\% increase in unique words (up to a maximum of 42\%) compared to those from random selection. Outputs selected using ART\_2gram and ART\_sBERT exhibit increases of 5.2\% and 4.8\%, respectively. 

The statistical significance of these improvements is confirmed by Wilcoxon Signed-Rank tests, all yielding very low p-values ($p < 10^{-4}$ for all methods). This result provides strong evidence for the effectiveness of diversity-based adaptive testing methods in promoting output diversity. It is worth noting that the considered outputs include reasoning steps (i.e., sequences of intermediate thought processes with the prefix ``Let's think step by step'' and before the final response), as well as the final responses required by the prompt.

\subsection{Cost Analysis (RQ3)}

\begin{figure}
    \centering
    \includegraphics[width=0.8\linewidth]{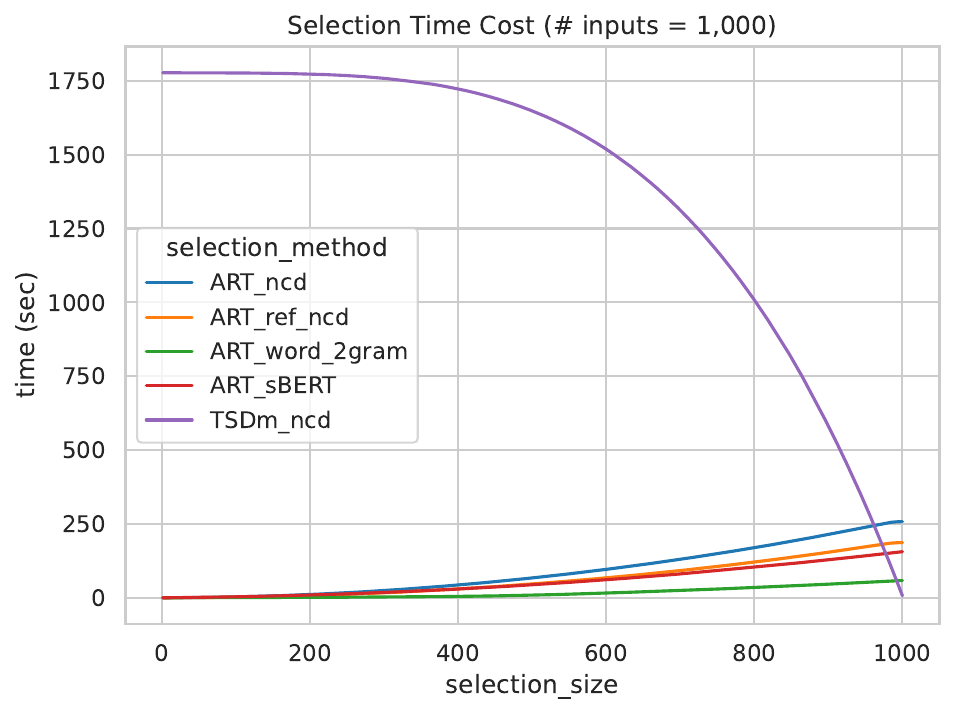}
    \caption{Time cost by different selection methods on different selection set size. Note that TSDm methods work by subtracting inputs, so the direction is the opposite to the ART methods.}
    \label{fig:RQ3_time_cost}
\end{figure}

To answer RQ3, we evaluate the computational overhead of selecting new test inputs using different methods. Figure~\ref{fig:RQ3_time_cost} shows the time required to calculate selections as the size of the selection set varies, using an initial input set of 1,000 inputs from the `news' task from P3 dataset, with an average input length of 257 characters. Note that the time required to calculate TSDm selections increases as the selection set size decreases. This is because TSDm calculates subsets by excluding inputs that minimally impact the the diameter of the previous subset.

Among ART distance variations, NCD shows the highest time cost, followed by sBERT-based distance (44 seconds) and ngram-based distance (9 seconds). We implemented cosine distance calculations for ngram and sBERT embeddings as parallelized matrix operations on the GPU, whereas NCD calculations are currently performed sequentially in a single-threaded manner. 

We observe that ART methods are significantly more efficient than TSDm when the initial pool size becomes substantially large ($N > 500$), as the time cost of TSDm scales with the size of the initial input set, whereas the time cost of ART is bounded by the size of the resulting selection set. For instance, selecting half of the inputs with the TSDm method takes 1,648 seconds, whereas ART\_NCD method requires only 67 seconds.
In a realistic test selection scenario, where a large number of user inputs are collected through application monitoring, TSDm can become easily impractical. However, given limited resources for executing inputs and labelling test outcomes, it is reasonable to assume that the target test set size will be kept relatively small. Therefore, ART remains a viable option as its computational cost is unaffected as long as the target selection set size is fixed.

We also compare the time cost of ART\_NCD when using the full reference set (i.e., all selected inputs so far) versus the selective reference set. For a selection size of 500 inputs (half the initial set), ART\_NCD with the selective reference set takes 47 seconds, compared to 67 seconds with the full reference set. This reduction is by design, as the selective reference set strategy decreases the size of the reference set, thereby reducing the number of required pairwise distance computations.

\section{Discussion}

\subsection{Integration with test input generators}

We focus on the selection scenario because it allows for practical evaluation, relying on labelled benchmarks with established ground truth. However, our adaptive testing method is not limited to this context and can be extended by incorporating a test input generator that dynamically produces new inputs. In this setup, the generator would iteratively create candidate inputs rather than sampling from an existing pool.

Emerging tools for testing prompt templates~\cite{promptfoo, deepeval} already support input generation using dedicated LLM agents. These agents can offer significant flexibility by allowing the generation process to be \textit{nudged} toward specific conditions. Building on this capability, a custom input generator could be developed to further refine the testing process. For instance, it could be programmed to produce inputs dissimilar to passing ones but similar to failing ones, enhancing the likelihood of detecting failures. This approach represents a promising avenue for future research, potentially enabling more effective failure detection in LLM-based applications and broadening the scope of our diversity-based adaptive testing method.

\subsection{Adaptive choice of effective distance metric}
\label{sec:discussion_distance_prediction}
Our empirical evaluation highlights that the effectiveness of diversity-based adaptive testing depends heavily on the choice of distance metric. For certain tasks, NCD significantly outperforms other metrics, while others fail to uncover failures and even perform worse than random baselines. This variation appears linked to task-specific characteristics---some tasks depend on syntactic distinctions, while others prioritize semantic meaning---and the data distribution within the initial test suite.

If this assumption holds, future research could focus on developing a method to predict the most effective distance metric for a given task and test suite. Such a method might analyze the embeddings of the initial test inputs, the distribution of distances among them, and\slash or the prompt template itself, aiming to identify patterns that reveal which embedding or distance metric that can best capture meaningful differences between test inputs. This approach could enhance the adaptability and overall effectiveness of diversity-based testing for LLM applications.

\subsection{Multi-modal inputs}
While most current LLM applications focus on processing textual inputs, there is increasing interest in multimodal LLMs and their diverse use cases. Notably, our diversity-based approach is not limited to text; it can be extended to any data type, provided suitable distance metrics are applied.

Normalized Compression Distance (NCD) is particularly promising in this context, as compression algorithms are inherently adaptable to various data types. For instance, in multimodal LLM applications where inputs combine text and images, NCD can be calculated by compressing the raw bytes of the input data. Furthermore, combining specialized compressors for each data type, e.g. text and images, could offer an even more precise measure of information similarity. This flexibility positions our approach as potentially applicable to evolving multimodal LLM-based systems.

\section{Conclusion}

We propose a diversity-based adaptive testing method for LLM applications, inspired by Adaptive Random Testing (ART) for conventional software. Our approach reduces the time and effort needed to uncover failures by prioritizing test inputs that are diverse from previously tested inputs.

We evaluate this method using two LLM evaluation datasets, comparing the performance of various distance metrics and selection strategies. The results show that our diversity-based approach can accelerate failure detection, enhance output diversity, while maintaining reasonable computational efficiency when selecting a fixed number of test inputs for manual review. This fully black-box method provides a practical, cost-effective solution for developers, improving test suite quality and streamlining the testing process.

\bibliographystyle{IEEEtran}
\bibliography{ref}

\end{document}